\documentclass{modelica}
\hypersetup{%
  pdftitle  = {Rumoca: Modelica as a Universal Algebraic Frontend via a Rust-Native Compiler},
  pdfauthor = {Micah K. Condie, Abigaile Woodbury, Thomas Meschede, Joel A.E. Andersson, James M. Goppert},
  pdfsubject = {American Modelica Conference 2026},
  pdfkeywords = {Modelica, compiler, Rust, WebAssembly, CasADi, JAX, Julia, benchmarking, AI-assisted development},
  colorlinks,
  linkcolor=black,
  urlcolor=black,
  citecolor=black,
  pdfpagelayout = SinglePage,
  pdfcreator = pdflatex,
  pdfproducer = pdflatex}

\usepackage[utf8]{inputenc}

\begin{document}
\thispagestyle{empty}

\title{Rumoca: Modelica as a Universal Algebraic Frontend via a Rust-Native Compiler}

\author[1]{Micah K. Condie}
\author[1]{Abigaile Woodbury}
\author[2]{Thomas Meschede}
\author[3]{Joel A. E. Andersson}
\author[1]{James M. Goppert}

\affil[1]{School of Aeronautics and Astronautics, Purdue University, USA, {\small\texttt{\{condiem,awoodbu,jgoppert\}@purdue.edu}}}
\affil[2]{Xyntopia LLC, USA, {\small\texttt{thomasm@xyntopia.com}}}
\affil[3]{FMIOPT AS, Norway, {\small\texttt{joel@fmiopt.com}}}

\maketitle\thispagestyle{empty}

\abstract{%
Modelica is a well established cyber-physical modeling language, but
many modern engineering workflows, such as optimization,
differentiable simulation, scientific machine learning, and system analysis, make use of other environments such as CasADi, JAX, and Julia. Existing infrastructure for Modelica toolchains
does not target these languages directly, meaning models must typically be rewritten, or lose important information in the interface. This paper presents Rumoca, a Rust-native
Modelica compiler that turns Modelica into a universal algebraic
frontend for a variety of tools. Rumoca is organized as a sequence of
explicit phase boundaries from parsing through Differential-Algebraic Equation (DAE) construction and
template-driven code generation, with a native Rust simulation
backend that also supports real-time, software-in-the-loop execution.
We report quantitative coverage of the Modelica Standard Library
across pipeline phases, together with compile-time and
simulation-time benchmarks against an open-source reference. The
full compiler ships as a VS Code extension and runs in the browser via WebAssembly, enabling
zero-install playgrounds and self-contained HTML simulators. End-to-end case studies demonstrate realtime
software-in-the-loop control of a quadrotor model and
deployment of a single Modelica source across multiple
algebraic backends.
}

\noindent\emph{Keywords: Modelica compiler, code generation, Rust, WebAssembly, FMI, CasADi, differentiable simulation, MSL benchmarking}

\section{Introduction}
\label{sec:intro}

The Modelica language is widely used for declarative, equation-based modeling
of cyber-physical systems. The mature
Modelica toolchain, led on the open-source side by OpenModelica
\cite{openmodelica} and complemented by industrial environments
such as Dymola, is highly effective at simulation, FMU export,
and graphical authoring. However, a growing set of engineering
workflows live downstream of simulation. Optimization,
differentiable simulation, scientific machine learning, and system analysis are all workflows typically built on symbolic and numerical frameworks such as CasADi \cite{Andersson2019CasADi}, JAX, and SymPy. Existing Modelica tools do not target these languages directly. In current practice, an engineer working across these ecosystems may end up writing the same model several times. For example, once in Modelica for simulation, again in CasADi for optimization, and a third time in SymPy for analysis, with no inherent guarantee that the three implementations of the model are the same. Moreover, these frameworks' modeling syntax is often harder to learn than Modelica's, since the model must be assembled imperatively in a general-purpose host language rather than declared as equations in a form close to textbook notation. Engineers who would benefit from Modelica's modeling discipline are often forced to use general-purpose languages or closed environments such as Simulink, where the tooling ecosystem is broader even when the modeling abstractions are weaker.

This paper presents Rumoca, a Rust-native Modelica compiler
that turns Modelica into a universal algebraic frontend for a
variety of tools. Rumoca preserves Modelica as a dedicated
declarative source language and exposes its
intermediate representation to template-driven backends that
emit CasADi, SymPy, FMI, JavaScript, JAX, Julia, ONNX, or embedded C, as well as
to a native-Rust simulation runtime with exact
automatic-differentiation Jacobians and software-in-the-loop (SIL)
support.

The remainder of this paper is organized as follows:

\begin{itemize}
  \item Section 2 reviews background on Modelica, FMI,
    and the Rumoca project.
  \item Section~\ref{sec:architecture} describes the Rumoca compiler
    architecture: a pipeline of explicit phase boundaries with typed
    intermediate representations, template-driven code generation to
    multiple algebraic backends, and a native Rust simulation backend
    with real-time SIL support, all exposed through
    unified command-line interface (CLI), language service protocol (LSP), WebAssembly, and Python frontends.
  \item Section 4 presents a quantitative Modelica Standard Library (MSL) parity
    benchmark across pipeline phases, together with compile- and
    simulation-time microbenchmarks against a reference open-source
    compiler.
  \item Section 5 describes Rumoca's WebAssembly-native
    deployment story, including self-contained HTML simulators that
    compile a Modelica model into a single shareable artifact
    runnable in any modern browser.
  \item Section 6 presents end-to-end case studies in which
    a single Modelica model is used as a realtime SIL
    plant for an external autopilot and as a source for multiple
    algebraic deployments.
\end{itemize}


\section{Background}
\label{sec:background}

\subsection{Modelica}

Modelica~\cite{MLSv34} is an
equation-based, object-oriented modeling language for
cyber-physical systems. Unlike procedural modeling approaches, Modelica models are expressed declaratively, with computational causality determined automatically by the compiler. This enables reusable component libraries and acausal multi-domain modeling. In procedural and signal-flow environments, computational causality is specified explicitly: block-diagram tools such as Simulink define directed computations between inputs and outputs, and hand-written Ordinary Differential Equations (ODE) implementations similarly require the modeler to choose which variables are solved for directly. In contrast, a Modelica equation such as $m\dot{v} = -kx - cv$ specifies a physical relationship without prescribing a computational direction. The compiler determines the required causality during transition.

The modeler can still fix a direction when one is wanted.
Top-level signals can be marked as \texttt{input} or
\texttt{output} for FMI export, for control design, or for
connecting to a causal environment. The equations inside a
component do not have to commit to a direction. As a result, the
same resistor, mass, or pipe model can be reused in different
contexts. Component libraries follow the same discipline. Typed
\texttt{connector} and \texttt{connect} equations let a library
author publish reusable parts. The language enforces the
interconnection rules so each user does not have to reimplement
them. The MSL ships several thousand
such components across electrical, mechanical, thermal, fluid,
and control domains.

Modelica is also a plain-text language. A model is a file. It can
be read, diffed, version-controlled, and generated by a script
in the same way as any other source code. This matters for large
language models. An LLM can read a Modelica file directly, reason
about its structure, and propose edits in the same medium. A
visual block diagram cannot be inspected this way. It has to be
serialized into some intermediate form first, if such a form
exists. As LLMs become a standard tool in engineering workflows,
the readability of the source language matters as much as its
expressive power. These properties make Modelica attractive even
when the downstream consumer is not a traditional simulator, such
as when the model feeds an optimizer, a learned controller, or a
symbolic analysis.

\autoref{lst:bouncing-ball} shows a complete Modelica model of a
bouncing ball. The model has a continuous dynamics
($\ddot{h} = -g$), an event condition ($h \le 0$ moving
downward), and a discrete reset that flips the velocity and
applies a coefficient of restitution. The compiler extracts the
state variables, detects the zero-crossing, and re-initializes
the integrator at each event.

\begin{lstlisting}[language=modelica, caption={A Modelica model of a bouncing ball. The \texttt{equation} section declares the continuous dynamics; the \texttt{when} clause declares an event-triggered discrete reset.}, label={lst:bouncing-ball}]
model BouncingBall
  parameter Real e = 0.8 "coefficient of restitution";
  parameter Real g = 9.81 "gravity";
  Real h(start = 1.0) "height";
  Real v(start = 0.0) "velocity";
equation
  der(h) = v;
  der(v) = -g;
  when h <= 0 and v < 0 then
    reinit(v, -e * pre(v));
  end when;
end BouncingBall;
\end{lstlisting}

\subsection{FMI}

The Functional Mock-up Interface (FMI)~\cite{fmi2022standard} is a
tool-independent standard for packaging simulation models as
Functional Mock-up Units (FMUs): zip archives containing an XML
interface description and a compiled implementation of the model's
dynamics. FMI defines two interaction modes. In \emph{Model
Exchange} the host provides the integrator and the FMU exposes
residuals, event indicators, and reset logic; in
\emph{Co-Simulation} the FMU embeds its own solver and is advanced
by the host over a communication step. The standard does not
specify a modeling language, but was developed in the Modelica
community and the two are commonly encountered together. Recent
work has extended FMI handling in
CasADi to support symbolic import, event dynamics, and analytic
sensitivities, narrowing the gap between FMI's traditional
black-box framing and the symbolic workflows targeted by Rumoca~\cite{andersson2024fmi, andersson2025events}.
Rumoca's FMI backend (\autoref{sec:architecture}) emits dual-mode
FMUs that declare both interfaces from a single Modelica source.

\subsection{Related Work}

The first commercial implementation
of a Modelica compiler is Dymola~\cite{dymola}, developed by Dynasim and now
owned by Dassault Systèmes. It remains the reference commercial
tool for simulation and FMU export. On the open-source side,
OpenModelica~\cite{openmodelica} is the most widely used compiler.
It provides simulation, FMU export, and graphical authoring
through OMEdit. A more recent open-source effort is
MARCO~\cite{Agosta2023Marco}, an LLVM/MLIR-based C++ compiler aimed at
high-performance simulation of large-scale models. These tools
all treat simulation as the primary deliverable. Rumoca instead
targets the symbolic and algebraic workflows that sit downstream
of simulation: optimization, differentiable simulation, scientific
machine learning, and symbolic analysis.

Another line of work worth mentioning is Julia-hosted modeling tools.
ModelingToolkit.jl~\cite{ma2021mtk} embeds an acausal,
equation-based modeling layer inside Julia. It exposes SciML
solvers and Julia's broader numerical ecosystem directly to the
modeler. Dyad~\cite{dyad}, released by JuliaHub in 2025 is a direct comparison to Modelica. It is a declarative, equation-based modeling language
with object-oriented composition, typed connectors, GUI authoring,
a standard component library, and FMI export. It covers
essentially the same design space as Modelica. The differences are
in the surrounding ecosystem and licensing. Dyad is released
under a source-available license rather than an open one. Its
compiler lowers Dyad source to ModelingToolkit.jl code, so models
ultimately run only inside the Julia/SciML stack. Its standard
component libraries are BSD3-licensed and are explicitly intended
to replace the Modelica Standard Library over time.

Rumoca takes a different stance. Rather than
introducing a new language tied to one runtime, it preserves Modelica
itself as an open, declarative frontend and emits the model into many
host languages, including ModelingToolkit.jl itself. The two approaches
trade off against each other: a clean-slate language like Dyad is free
to introduce new modeling features directly, whereas Rumoca, building on
the existing Modelica language, would need such features added to
Modelica itself. In exchange, Rumoca inherits Modelica's 25-year history
and large public corpus of open-source models, including the MSL. This is an
advantage that also benefits LLM workflows, since an LLM can be trained
on or grounded against that corpus.

\subsection{Evolution of Rumoca}

In our previous paper~\cite{condie2025rumoca} we introduced Rumoca
as a Rust-based translator from Modelica to algebraic modeling
languages, with initial CasADi and SymPy backends and a
Jinja-driven template mechanism for additional targets. That paper
positioned Rumoca as a translator, demonstrated it on three
example models, and outlined a path in which complex Modelica
models would first be lowered to flattened Modelica by an external compiler before being handed to Rumoca.

Since that publication, Rumoca has matured into a complete
Modelica compiler. This change in scope was made possible by AI co-development (see \autoref{sec:ai-development}). Even so, implementing full Modelica support takes time, and using flattened Modelica as a bridge between more mature compilers and Rumoca is still necessary in some cases. Rumoca now implements parsing, name resolution, type
checking, instantiation, flattening, DAE construction, structural
analysis with initialization-condition planning, and
template-driven code generation across explicit phase boundaries.
A native Rust simulation backend provides exact
automatic-differentiation Jacobians and solver fallbacks, the full
pipeline runs in the browser via WebAssembly, and the project
maintains a continuous-integration parity gate against the bulk of
the MSL. Rumoca is distributed through CLI, language-server,
Python, and VS Code interfaces. The remainder of this paper
describes the resulting system.

\section{Architecture, Features, and Development}
\label{sec:architecture}
Rumoca is designed such that each phase of the compiler has a
clear scope and produces an accessible artifact. Solvers, code
generators, and simulators all consume clear intermediate
representations, but none of these consumers is permitted to
influence the IRs. Rather, each IR is either transformed into
the next IR by a compiler phase, or consumed terminally by a
downstream tool. This clear separation of representations is
what aids the design of Rumoca as a Modelica frontend for many ecosystems rather than a simulator alone. Other than code generation, features of Rumoca include SIL and native Rust simulation with automatic differentiation. Rumoca can be used through the CLI, through a VS Code extension, and imported as a Python package (a pip-installable wheel) or npm package.

\subsection{Pipeline Overview}
\label{sec:pipeline}

Rumoca consists of two main peer activities, \emph{Compile} and \emph{Simulation}. Compile is responsible for parsing and building the various internal representations of the model, while Simulation takes the IR representations produced by the Compile API and applies the relevant solvers and IO tools required for simulation. Internally, every activity in Rumoca is implemented as a sequence of phases, each phase living in its own Rust crate and communicating only through explicit transitions. This organization, of clearly defining crate scope, mirrors that of the Rust compiler itself \cite{Rustc-dev-guide}, whose source tree is split across dozens of crates with strict phase boundaries. Compile and Simulation activities are exposed by the four entry points of Rumoca: the  command-line binary, an LSP server (used in the VS Code extension), Python bindings, and WebAssembly bindings. These will be described in more detail later.

\begin{figure}[htbp]
\centering
\includegraphics[width=\columnwidth]{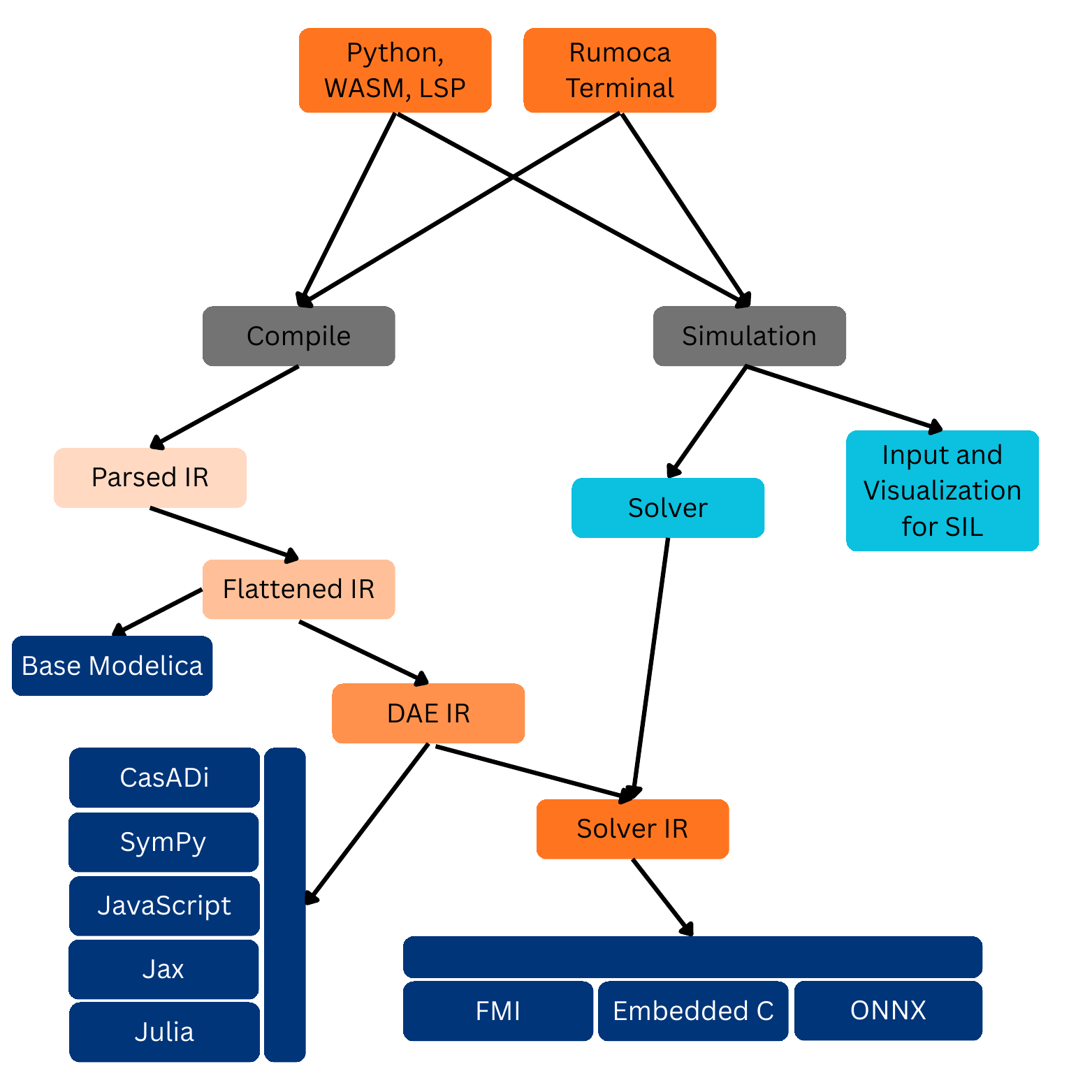}
\caption{Rumoca's architecture. Compile and Simulation are peer activities, each callable from any frontend. The IR's are used to generate the various backends. The Solver IR used for Rumoca's native simulation}
\label{fig:pipeline}
\end{figure}

\autoref{fig:pipeline} shows an overview of the Rumoca architecture, though in reality the implementation consists of 45 Rust crates. Within the compile activity, there are four stages of processing a Modelica model goes through, each represented by a unique IR. The first pass on the model is done by the Parol \cite{singer2025parol} parser. This produces the Parsed IR that still contains the native object-oriented structure of the model. The second phase then flattens this model, producing the Flattened IR, which has the object-oriented structure removed and the connect equations resolved. This bears the closest resemblance to traditional flattened Modelica. The next pass produces the DAE IR, which arranges the equations of the model in a hybrid differential-algebraic form as specified by the Modelica Language Specification Appendix~B. This includes variables being tagged by role (state, algebraic, discrete, parameter) and equations tagged by category (continuous, discrete update, event condition). A second pass reduces the system to index 1, the form in which all state derivatives can be recovered after at most one differentiation, while keeping it in the same DAE IR structure. A final pass produces the Solver IR, which represents the model lowered to a flat, register-based opcode stream with Jacobian-vector products precomputed by forward-mode automatic differentiation, ready for direct numerical evaluation by an interpreter or a JIT backend. This multi-pass discipline, where each phase is distinct and crosses a typed IR boundary, simplifies testing and AI-assisted contribution. It also opens the door to advanced diagnostics: when a simulation or optimization fails, the user can inspect each IR in turn---Parsed, Flattened, DAE, or Solver---to localize the problem to a specific compiler phase.

\subsection{Code Generation}
\label{sec:codegen}

Code generation is template-driven and can operate on any of the IR phases. Templates are processed by the Jinja2 template engine~\cite{jinja2}, a widely used text-templating system originating in the Python web-development ecosystem, where it is commonly used to generate HTML pages from dynamic data. The same mechanism applies cleanly to code generation: Jinja2 walks the prepared DAE and emits the model in the syntax specified by a particular template. Each backend is a separate template; Jinja2's control-flow constructs (loops, conditionals) handle the per-model variation within a template. New backends can be added by writing additional templates, using existing ones as examples. The templates currently shipped with Rumoca, in approximate order of maturity, are:

\begin{itemize}
  \item \textbf{CasADi} in both MX and SX
    expression flavours --- MX for matrix-valued DAGs of arbitrary
    CasADi nodes, SX for dense scalar-valued trees.
  \item \textbf{SymPy} for symbolic analysis and verification.
  \item \textbf{FMI 2.0 and FMI 3.0}, exported as dual-mode FMUs
    that declare both Model Exchange and Co-Simulation; Model
    Exchange is the accurate path, with Co-Simulation provided as
    a convenience wrapper for tools that only consume CoSim FMUs.
  \item \textbf{JavaScript} exported as self-contained js code, including solvers for inclusion in web-based frontends.
  \item \textbf{JAX} for differentiable simulation and machine learning.
  \item \textbf{Julia / ModelingToolkit} that works with SciML solvers.
  \item \textbf{ONNX} for export of model components into
    machine-learning runtimes.
  \item \textbf{Embedded C}, currently at an early stage: the
    template generates a self-contained C model suitable for
    integration into bare-metal or RTOS targets, and has been
    flown experimentally on a quadrotor through the CogniPilot
    Cerebri autopilot, but is not yet recommended for production
    use.
\end{itemize}

In addition to these consumer backends, Rumoca ships two other templates. The first produces flat Modelica, adhering to the proposed Base Modelica standard \cite{mcp0031}. The second re-emits the compiler's own DAE and flat-model IRs as Modelica source. These are not translation targets in the
conventional sense but inspection and traceability surfaces:
they let a user see exactly what the flattening or
DAE-formation phase has produced. The DAE re-emission is
particularly useful as an entry point for custom tools and
analyses, since it exposes Rumoca's canonical hybrid DAE in a
human- and machine-readable form that downstream consumers can parse without re-implementing any of the upstream compiler
phases. It also provides a natural handoff to other tools
expecting Base Modelica--style input.

\subsection{Simulation and SIL}
 Rumoca's simulation runtime consumes the Solver IR and ships
with two solver backends: a pure-Rust RK45 explicit integrator
for non-stiff problems, and an integration with Diffsol
\cite{Robinson2026Diffsol}, an open-source Rust ODE and
semi-explicit DAE solver from the University of Oxford, which
provides BDF and ESDIRK methods for stiff and implicit-DAE
problems. Initialization-condition planning runs ahead of
integration, and residual evaluation can be optionally JIT-compiled
through Cranelift \cite{cranelift} on native targets and through WebAssembly
in the browser. Jacobians and mass matrices are obtained by
exact forward-mode automatic differentiation of the DAE IR
rather than finite differences, giving the solver analytic
derivatives at every step. The runtime supports both batch
execution and a deterministic, step-locked realtime mode
suitable for SIL work, with named-signal I/O
serialized through a FlatBuffers codec \cite{flatbuffers} and transported over
UDP for desktop clients or WebSockets for browser viewers and
remote operators. Live operator input is supplied through
gamepad and keyboard backends. For visualization, simulations
can drive a Three.js scene, giving the operator a live 3D view
of the plant alongside the numeric feed.
 
\subsection{Additional Features}
\label{sec:session}

The language server, exposed through the \emph{Rumoca Modelica}
extension on the VS Code marketplace, makes the compiler's
internal information available at edit time. Beyond standard
LSP features --- realtime diagnostics, autocomplete, hover
information, go-to-definition, find-references, document
symbols, signature help, and formatting --- the extension
integrates directly with the compiler's simulation surface.
A user can edit a \code{.mo} file, trigger a one-click
simulation run, and receive a self-contained HTML report with
embedded plots, all without leaving the editor. The extension
also exposes a configurable visualization view system: in
addition to standard time- and X/Y-plot views, a 3D view
generates a Three.js scene script for the model that the user
can edit directly to reflect the model's physical state, giving
each Modelica model a customizable runtime visualization
without leaving the source-tree workflow.

\subsection{AI-Assisted Compiler Development}
\label{sec:ai-development}

Rumoca has been developed with substantial assistance from
large-language-model based coding agents. This has only worked under close
human direction and with project-specific guardrails. For semantic changes, we
require an explicit lookup of the relevant Modelica Language Specification
rule. We also require the agent to respect Rumoca's phase boundaries, identify
the owning compiler phase, and avoid fixing downstream consumers when the error
belongs in the compiler pipeline.

In practice, the useful workflow is to define the compiler contract first and
let the agent implement inside that boundary. For compiler bugs, we usually ask
the agent to reproduce one concrete failure, trace the affected symbol or
equation through the pipeline, inspect intermediate artifacts such as flattened
models or DAEs, and compare phase-level MSL outcomes. This helps distinguish a
real regression from a change that improves an earlier phase while exposing a
pre-existing limitation in a later one.

Despite these measures, the main limitation remains unresolved: agents still
struggle with root-cause analysis and frequently repair the visible symptom. A
simulation failure may lead to a solver-side workaround. A diagnostic failure
may lead to relaxed validation. A backend mismatch may lead to a template patch.
In each case, the actual bug may have been introduced earlier during name
resolution, instantiation, flattening, or DAE construction.

As a result, substantial manual analysis is still required. Human developers
must guide the investigation, inspect compiler artifacts, decide which phase
first introduced the wrong symbol, equation, or structural form, and reject
plausible but shallow fixes. We therefore treat AI output as a hypothesis, not a
fix. In our experience, AI accelerates implementation once the semantic target
is clear, but it does not replace human root-cause analysis, specification
judgment, or architectural review.

\section{Benchmarking and MSL Parity}
\label{sec:benchmarking}

To quantify how much of the Modelica language Rumoca handles in
practice, we benchmark the compiler against the Modelica Standard
Library (MSL) version 4.1.0~\cite{msl410}, the canonical reference
suite for Modelica tools. Rather than report a single pass/fail
number, we measure two things: how far each model advances through the
pipeline described in \autoref{sec:architecture} (a reachability
funnel that separates front-end coverage from DAE construction, solver
lowering, initialization, and simulation), and, for the models that
simulate to completion, how closely their trajectories agree with the
OpenModelica compiler (OMC).

\subsection{Methodology}
\label{sec:msl-method}

The benchmark set is the collection of root example models in MSL---
every non-partial class of the form
\texttt{Modelica.<library>.Examples.<...>} with no unbound top-level
inputs. This yields 566 standalone-simulatable models spanning the
MSL sublibraries. All OMC references were produced with OpenModelica
1.26.7.

\paragraph{Reachability funnel.}
Each model is run through the full Rumoca pipeline, and we record the
last phase it reaches: \emph{Parse} (Parsed IR), \emph{Flatten}
(Flattened IR), \emph{DAE} (DAE IR), \emph{Solve} (Solver IR),
\emph{IC} (consistent initialization), and \emph{Sim} (integration to
the model's stop time through the embedded Diffsol suite in its
\texttt{auto} mode, which tries BDF first and falls back through
RK-family methods on stiff or discontinuous problems). A model counts
as passing a phase if the pipeline gets past that phase without
failing; the phases are nested by construction, so the reported rates
are non-increasing from left to right across all packages. \emph{Sim}
here measures only that a model integrates to completion, not that it
agrees with OMC; agreement is measured separately below. Rumoca
integrates at solver tolerances
$\mathrm{rtol}=\mathrm{atol}=10^{-6}$.

\paragraph{Matched stop time.}
Both tools integrate each model to its own
\texttt{experiment(StopTime=$\cdot$)} annotation: Rumoca reads the
annotation directly, and OMC is driven with
\texttt{-{}-use-experiment-stop-time} so it uses the same horizon. This
makes the integration spans match per model, which is a precondition
for a meaningful simulation comparison.

\paragraph{Trace agreement.}
For the models that both tools simulate to completion, parity is
measured per output channel with a normalized, bounded error. Channels
are variables matched by exact name; each channel is compared on the
union of both tools' sample times over the overlap window, with linear
interpolation for continuous channels and zero-order hold for
discrete/clocked channels. For a channel, let $\bar{e}$ be the
time-averaged absolute deviation between the Rumoca and OMC traces
(trapezoidal rule), normalized by a robust scale $s$ equal to the
$5$th--$95$th percentile span of that variable's OMC trace; the
channel score $e = \bar{e}/s$ is mapped into $[0,1)$ by
$\hat{e} = e/(1+e)$, so a single diverging or near-flat channel
saturates near $1$ rather than dominating the aggregate. A channel is
in \emph{high} agreement if $\hat{e} \le 0.05$, \emph{near} agreement
if $\hat{e} \le 0.20$, and \emph{deviating} otherwise. A model is
\emph{agreeing} if its channel distribution falls in either a
high-agreement band (at least $80\%$ of channels high and at most
$1\%$ deviating) or a near-agreement band (at least $90\%$ high-or-near
and at most $10\%$ deviating). Of the 151 models that both tools
simulate and that produce comparable traces, 102 ($\approx 67\%$) are
agreeing; the remaining third deviate. The agreeing set is the basis
for the timing comparison in \autoref{sec:speed}.

\subsection{Coverage Results}

\autoref{tab:msl_parity} reports per-package and overall pass rates at
each phase. All 566 models parse cleanly (100\% for every sublibrary),
so we omit the Parse column and begin at Flatten. Flattening succeeds
for 99\% of the suite; the only sublibraries below 100\% are
\texttt{Electrical.Polyphase}, \texttt{Electrical.Spice3},
\texttt{Magnetic.FluxTubes}, and \texttt{Utilities}.

DAE construction succeeds for 73\% of the suite. The losses are
concentrated in connection- and constraint-heavy libraries: the
quasi-static electrical machines and polyphase libraries (0\%), the
magnetic wave libraries (\texttt{FundamentalWave} 4\%,
\texttt{QuasiStatic.FundamentalWave} 10\%), \texttt{Fluid} (35\%), and
the large coupled-constraint systems of \texttt{Mechanics.MultiBody}
(55\%); \texttt{StateGraph} (0\%) is gated by state-machine semantics
not yet lowered to a DAE. \texttt{Media} now constructs a DAE for
68\% of its examples.

Solver lowering (60\%) and initialization (41\%) are the next two
attrition steps. \texttt{Electrical.Machines}, for example, lowers a
Solver IR for 53\% of its models but loses all of them at
initialization (53\% Solve $\rightarrow$ 0\% IC), and
\texttt{Magnetic.FluxTubes} drops from 85\% Solve to 15\% IC.

End-to-end, 28\% of the suite simulates to its stop time, and---as
noted in \autoref{sec:msl-method}---about two-thirds of the simulated
models then agree with OMC. There are thus two distinct gaps: a
\emph{reachability} gap (only 28\% reach a completed simulation) and an
\emph{agreement} gap (of those, $\approx 33\%$ deviate from OMC).
Several libraries are clean end-to-end:
\texttt{ComplexBlocks} and the \texttt{Math} examples reach 100\%,
\texttt{Electrical.Digital} 96\%, and \texttt{Clocked} 88\%. By
contrast, several switched-electronics and thermal libraries reach
high DAE and solver coverage but simulate few or no models
(\texttt{Electrical.PowerConverters}: 76\% Solve, 66\% IC, but 0\%
Sim), indicating gaps in event handling and stiff-solver coverage
rather than missing language features. Compile-side phases are fast
(parse, flatten, and DAE construction average well under a second per
model); wall-clock time is dominated by the solve and simulation
phases.

\begin{table*}[t]
  \centering
  \caption{MSL 4.1.0 pass rates by package and pipeline phase.
  Scope: $N = 566$ root \texttt{Modelica.*.Examples.*} models. All
  models pass Parse at 100\%, so that column is omitted. A model
  passes a phase (Flatten, DAE, Solve, IC, Sim) if the pipeline
  reached past it; \emph{Sim} means the model integrated to its
  \texttt{experiment} stop time, not that it agrees with OMC (trace
  agreement is reported separately in \autoref{sec:msl-method}).
  Reported rates are non-increasing left-to-right per row.}
  \label{tab:msl_parity}
  \footnotesize
  \begin{tabular}{lrrrrrr}
    \toprule
    MSL Package & $n$ & Flatten & DAE & Solve & IC & Sim \\
    \midrule
    Blocks                               & 32  & 100\% & 78\%  & 69\%  & 63\%  & 63\%  \\
    Clocked                              & 76  & 100\% & 96\%  & 96\%  & 88\%  & 88\%  \\
    ComplexBlocks                        & 2   & 100\% & 100\% & 100\% & 100\% & 100\% \\
    Electrical.Analog                    & 65  & 100\% & 95\%  & 83\%  & 58\%  & 32\%  \\
    Electrical.Batteries                 & 8   & 100\% & 50\%  & 13\%  & 13\%  & 0\%   \\
    Electrical.Digital                   & 23  & 100\% & 96\%  & 96\%  & 96\%  & 96\%  \\
    Electrical.Machines                  & 43  & 100\% & 63\%  & 53\%  & 0\%   & 0\%   \\
    Electrical.Polyphase                 & 5   & 80\%  & 80\%  & 80\%  & 20\%  & 20\%  \\
    Electrical.PowerConverters           & 59  & 100\% & 98\%  & 76\%  & 66\%  & 0\%   \\
    Electrical.QuasiStatic.Machines      & 1   & 100\% & 0\%   & 0\%   & 0\%   & 0\%   \\
    Electrical.QuasiStatic.Polyphase     & 4   & 100\% & 0\%   & 0\%   & 0\%   & 0\%   \\
    Electrical.QuasiStatic.SinglePhase   & 6   & 100\% & 100\% & 83\%  & 0\%   & 0\%   \\
    Electrical.Spice3                    & 14  & 93\%  & 79\%  & 79\%  & 7\%   & 0\%   \\
    Fluid                                & 23  & 100\% & 35\%  & 0\%   & 0\%   & 0\%   \\
    Magnetic.FluxTubes                   & 20  & 90\%  & 90\%  & 85\%  & 15\%  & 0\%   \\
    Magnetic.FundamentalWave             & 27  & 100\% & 4\%   & 4\%   & 0\%   & 0\%   \\
    Magnetic.QuasiStatic.FluxTubes       & 9   & 100\% & 100\% & 100\% & 0\%   & 0\%   \\
    Magnetic.QuasiStatic.FundamentalWave & 21  & 100\% & 10\%  & 0\%   & 0\%   & 0\%   \\
    Math.FastFourierTransform            & 2   & 100\% & 100\% & 100\% & 100\% & 100\% \\
    Math.Nonlinear                       & 1   & 100\% & 100\% & 100\% & 100\% & 100\% \\
    Math.Random                          & 1   & 100\% & 100\% & 100\% & 100\% & 100\% \\
    Mechanics.MultiBody                  & 42  & 100\% & 55\%  & 0\%   & 0\%   & 0\%   \\
    Mechanics.Rotational                 & 17  & 100\% & 100\% & 100\% & 41\%  & 18\%  \\
    Mechanics.Translational              & 16  & 100\% & 100\% & 100\% & 81\%  & 69\%  \\
    Media                                & 22  & 100\% & 68\%  & 36\%  & 27\%  & 27\%  \\
    Media.Incompressible                 & 1   & 100\% & 100\% & 0\%   & 0\%   & 0\%   \\
    StateGraph                           & 7   & 100\% & 0\%   & 0\%   & 0\%   & 0\%   \\
    Thermal.FluidHeatFlow                & 12  & 100\% & 17\%  & 17\%  & 8\%   & 0\%   \\
    Thermal.HeatTransfer                 & 4   & 100\% & 100\% & 100\% & 75\%  & 25\%  \\
    Utilities                            & 3   & 67\%  & 67\%  & 67\%  & 67\%  & 67\%  \\
    \midrule
    Overall                              & 566 & 99\%  & 73\%  & 60\%  & 41\%  & 28\%  \\
    \bottomrule
  \end{tabular}
\end{table*}

\subsection{Compile and Simulation Time}
\label{sec:speed}

We complement the coverage results with a wall-clock comparison
against OpenModelica (OMC), reporting \emph{compilation} and
\emph{simulation} time separately rather than as a single combined
number. The two tools have opposite performance profiles: Rumoca's
advantage is concentrated in compilation, while its interpreted
integration loop is slower than OMC's native code on large systems, so
a combined figure would hide the crossover. All timings were collected
on a single workstation running Ubuntu~24.04 with an AMD Ryzen~9
5950X (16 cores, 32 threads) and 64\,GB of RAM.

The harness runs a pool of 14 worker processes, each pinned to a
dedicated physical core, that draw models from the 100-model agreeing
set (the models that both tools simulate to their
\texttt{experiment} stop time and on which the traces agree within the
bands of \autoref{sec:msl-method}, and for which both tools record
valid timings) until the set is exhausted. For each model we measure
the time to build and simulate it from scratch; a model exceeding a
10\,s timeout is killed and its worker restarted. Each Rumoca worker
retains a persistent in-process cache, so intermediate representations
of shared MSL components computed for one model may be reused by later
models in the same worker. The OMC side is driven through its
persistent ZMQ session, so it is likewise kept warm across the run.
We separate three Rumoca stages---front-end through DAE, Solver-IR
lowering plus Cranelift JIT build, and the integration loop---and
group the first two as \emph{compilation} (build-to-runnable) and the
third as \emph{simulation}. For OMC we use the self-reported
$\texttt{timeTotal}-\texttt{timeSimulation}$ as compilation (frontend,
backend, SimCode, C-code generation, and gcc) and
$\texttt{timeSimulation}$ as simulation. We report median per-model
time within scalar-equation-count groups, and a throughput (sum/sum)
ratio across the whole set.

\autoref{tab:speed} reports the results in three panels.
\emph{Compilation} (panel b): Rumoca is faster in every bin
($2.6$--$5.2\times$; throughput $4.18\times$, front-end-only median
$5.25\times$), because it emits native code through a Cranelift
JIT~\cite{cranelift} rather than generating C and invoking gcc.
\emph{Simulation} (panel c) crosses over with system size: Rumoca
leads on small models ($6.7\times$ at 1--9 equations), reaches parity
near 25--49 equations, and then falls behind sharply
($0.29\times$, $0.05\times$, $0.01\times$ in the top three bins;
throughput $0.22\times$). OMC's integration is nearly flat
($\approx 0.02$--$0.04$\,s) because it runs native compiled C, while
Rumoca's interpreted Solver-IR runtime grows steeply with size.
\emph{Total} (panel a): end to end, Rumoca is faster on all but the
very largest systems---median per-model speedup $4.79\times$,
throughput $3.01\times$---and only the 250+ bin (two models) tips to
OMC, at $0.54\times$. The breakdown localizes the gain: the
speedup is a compile-time effect, and Rumoca's numerical integration is
in fact slower than OMC's on large systems. 

These numbers should be read as an efficiency comparison, not a
like-for-like one. The comparison is restricted to the subset of
models that both tools simulate successfully and on which they agree,
which by construction excludes the harder cases where OMC's index
reduction, tearing, and broader stiff- and event-handling succeed and
Rumoca does not; this easy-and-agreeing subset
plausibly favors Rumoca, and OMC remains the more robust and capable
simulator. Several further issues temper the absolute numbers. The
warm, core-pinned worker pool lets Rumoca reuse cached IR for shared
MSL components across models, and although OMC is driven through its
persistent ZMQ session, we have not verified that it caches MSL
components equivalently, so part of the measured compile gap may
reflect caching asymmetry. The agreeing population skews small (62 of
the 100 models have $\le 24$ equations), so per-model median speedups
are small-model-weighted; we report the throughput ratio alongside
every median for this reason, and the largest ($250+$) bin contains
only two models. Finally, each model is timed once on a single host
with no variance bars; a production claim should repeat the run and
report variance, as sub-second compile times are sensitive to
scheduling and thermal noise. We therefore present these results as
evidence that the Rumoca pipeline is fast enough to be practical, not
as a claim of general superiority over a production Modelica tool.

\begin{table}[t]
  \centering
  \caption{Wall-clock time, Rumoca vs.\ OpenModelica (OMC), on the 100
  models that both tools simulate to their \texttt{experiment} stop
  time and agree within the parity thresholds of
  \autoref{sec:msl-method}, grouped by scalar-equation count.
  \emph{Compilation} is build-to-runnable
  ($\texttt{timeTotal}-\texttt{timeSimulation}$ for OMC, including
  gcc); \emph{Simulation} is the integration loop only. Per-bin
  Rumoca/OMC columns are median seconds and Speedup is the per-bin
  median of per-model ratios; the \emph{All} row reports the
  throughput (sum/sum) ratio over all 100 models, which need not equal
  any median.}
  \label{tab:speed}
  \footnotesize
  \begin{tabular}{lrrrr}
    \toprule
    Scalar eqns & Models & Rumoca (s) & OMC (s) & Speedup ($\times$) \\
    \midrule
    \multicolumn{5}{l}{\emph{(a) Total: model $\rightarrow$ simulated result}}\\
    1--9     & 26  & 0.46 & 2.40 & 5.20 \\
    10--24   & 36  & 0.49 & 2.44 & 5.03 \\
    25--49   & 11  & 0.54 & 2.48 & 4.55 \\
    50--99   & 14  & 0.80 & 2.68 & 3.37 \\
    100--249 & 11  & 1.33 & 2.75 & 2.07 \\
    250+     & 2   & 6.05 & 3.29 & 0.54 \\
    All      & 100 & 84.9 & 255.9 & 3.01 \\
    \addlinespace
    \multicolumn{5}{l}{\emph{(b) Compilation: build-to-runnable}}\\
    1--9     & 26  & 0.46 & 2.38 & 5.18 \\
    10--24   & 36  & 0.48 & 2.42 & 5.10 \\
    25--49   & 11  & 0.50 & 2.45 & 4.88 \\
    50--99   & 14  & 0.55 & 2.58 & 4.72 \\
    100--249 & 11  & 0.97 & 2.73 & 2.81 \\
    250+     & 2   & 1.24 & 3.26 & 2.64 \\
    All      & 100 & 59.8 & 250.3 & 4.18 \\
    \addlinespace
    \multicolumn{5}{l}{\emph{(c) Simulation: integration only}}\\
    1--9     & 26  & 0.0028 & 0.0190 & 6.71 \\
    10--24   & 36  & 0.0100 & 0.0205 & 2.06 \\
    25--49   & 11  & 0.0255 & 0.0252 & 0.99 \\
    50--99   & 14  & 0.1376 & 0.0399 & 0.29 \\
    100--249 & 11  & 0.4889 & 0.0240 & 0.05 \\
    250+     & 2   & 4.8140 & 0.0292 & 0.01 \\
    All      & 100 & 25.1   & 5.6    & 0.22 \\
    \bottomrule
  \end{tabular}
\end{table}
\section{Browser-Native Compilation via WebAssembly}
\label{sec:wasm}

A central goal of Rumoca is to make Modelica modeling easy to use. The compiler infrastructure described in
\autoref{sec:architecture} is built so that the
WASM bindings crate exposes the compiler's
API as a JavaScript module. Two distinct
deployment patterns fall out of this design.

\subsection{Compiler-in-Browser}
\label{sec:compiler-in-browser}

\begin{figure}[htbp]
\centering
\includegraphics[width=\columnwidth]{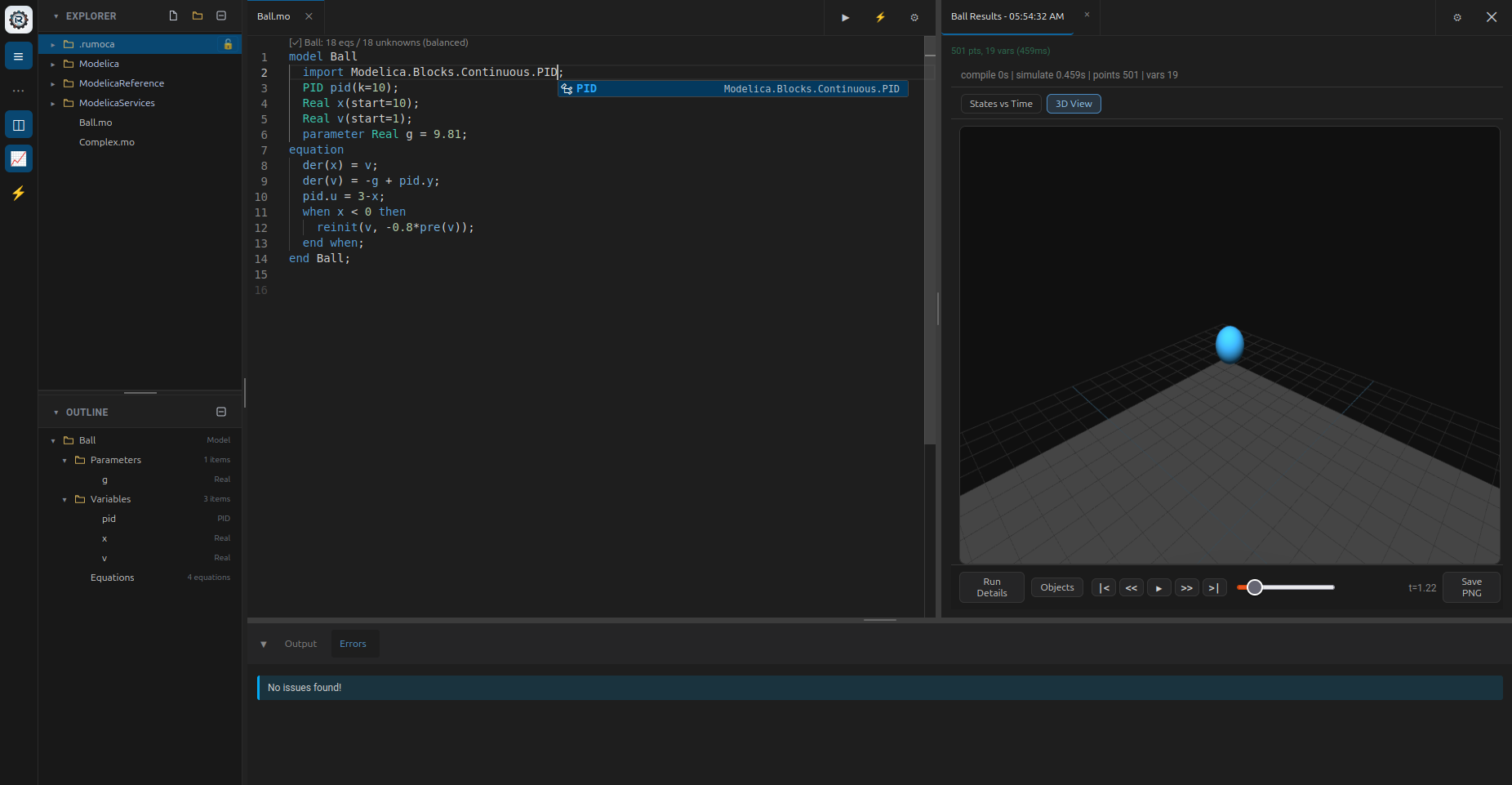}
\caption{The Rumoca in-browser playground. The compiler runs entirely client-side via WebAssembly, with parsing, diagnostics, and DAE inspection driven by the same \code{Session} API used by the desktop CLI and language server.}
\label{fig:playground}
\end{figure}

The first pattern runs the compiler itself client-side. A
hosted playground at \url{https://rumoca.cognipilot.org/} (\autoref{fig:playground}) loads
the WASM build of \code{rumoca-bind-wasm} in the user's browser
and uses it to drive an editor with realtime parsing,
diagnostics, and DAE introspection on every keystroke. The
implementation stack matches the desktop language server
described in \autoref{sec:session}: the same \code{Session},
the same diagnostic types, the same backend templates.

This is unusual for Modelica tooling. Browser-based Modelica
environments typically run the compiler on a remote server and
ship results to the client; the user's models leave their
machine, the server bears the operating cost, and editor latency
is bounded by network round-trips. By contrast, the Rumoca
playground requires no backend, no authentication, no
per-session quota. Models stay on the user's machine. The
operator of the playground pays only for static asset hosting.
This is particularly valuable to researchers, students, and engineers evaluating Modelica without committing to a desktop installation, lowering the barrier to first contact.

\subsection{Standalone HTML Simulators}
\label{sec:standalone-html}

\begin{figure}[htbp]
    \centering
    \includegraphics[width=1\linewidth]{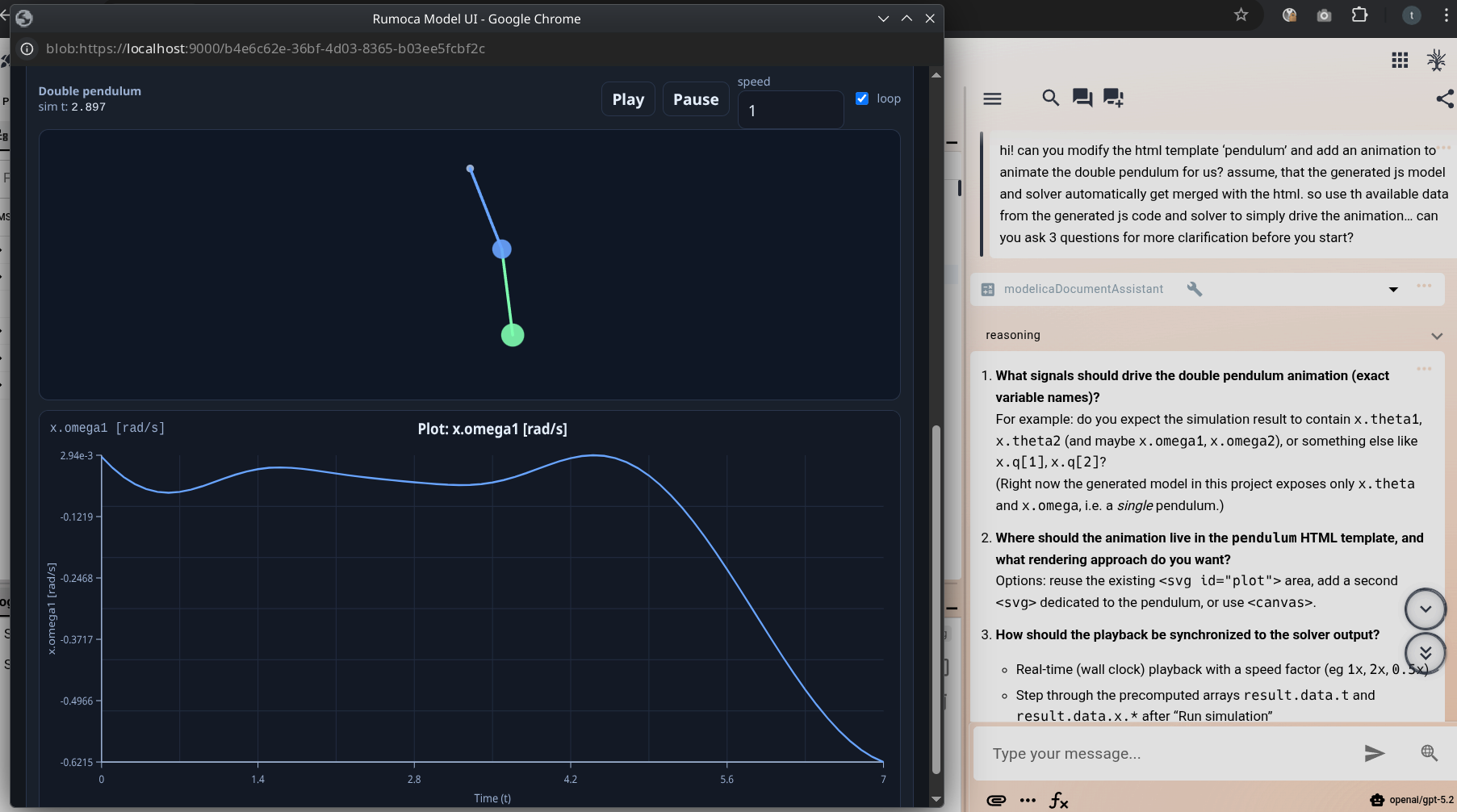}
    \caption{Modelica simulations can be exported into fully self-contained, pure HTML/js based and customizable files using rumoca with its js based templates.}
    \label{fig:js_sim}
\end{figure}

Rumoca includes JavaScript templates that compile Modelica models into pure, self-contained JavaScript code, which can be embedded directly in an HTML file (\autoref{fig:js_sim}). This enables lightweight, browser-based simulators and makes Rumoca suitable for applications such as digital twins, product model exchange, and interactive simulation demos on web pages.

The js-template-based approach exposes the solver and code-generation logic in a form that is easy for AI coding agents to inspect, modify, and extend. Existing examples \cite{taskyon2026} demonstrate how compilation targets can be switched quickly between different architectures and programming languages or build rich interfaces on top of detailed simulation models.

%

%
%
%

\section{Case Studies}
\label{sec:casestudies}
This section presents two end-to-end case studies, each taking a single
Modelica source through the Rumoca toolchain. \ref{sec:sil} drives a
21-state quadrotor as a realtime SIL
plant for an external autopilot, exercising the native Rust runtime.
\ref{sec:multibackend} takes a small planar quadrotor and emits it into
four algebraic ecosystems from one source, exercising the code
generators. Together they support the paper's thesis: Modelica as a
universal algebraic frontend, authored once and consumed by downstream
optimization, symbolic analysis, differentiable simulation, and
tool-independent deployment in their native idioms.

\subsection{Realtime Software-in-the-Loop with Gamepad Control}
\label{sec:sil}

\begin{figure}[htbp]
\centering
\includegraphics[width=\columnwidth]{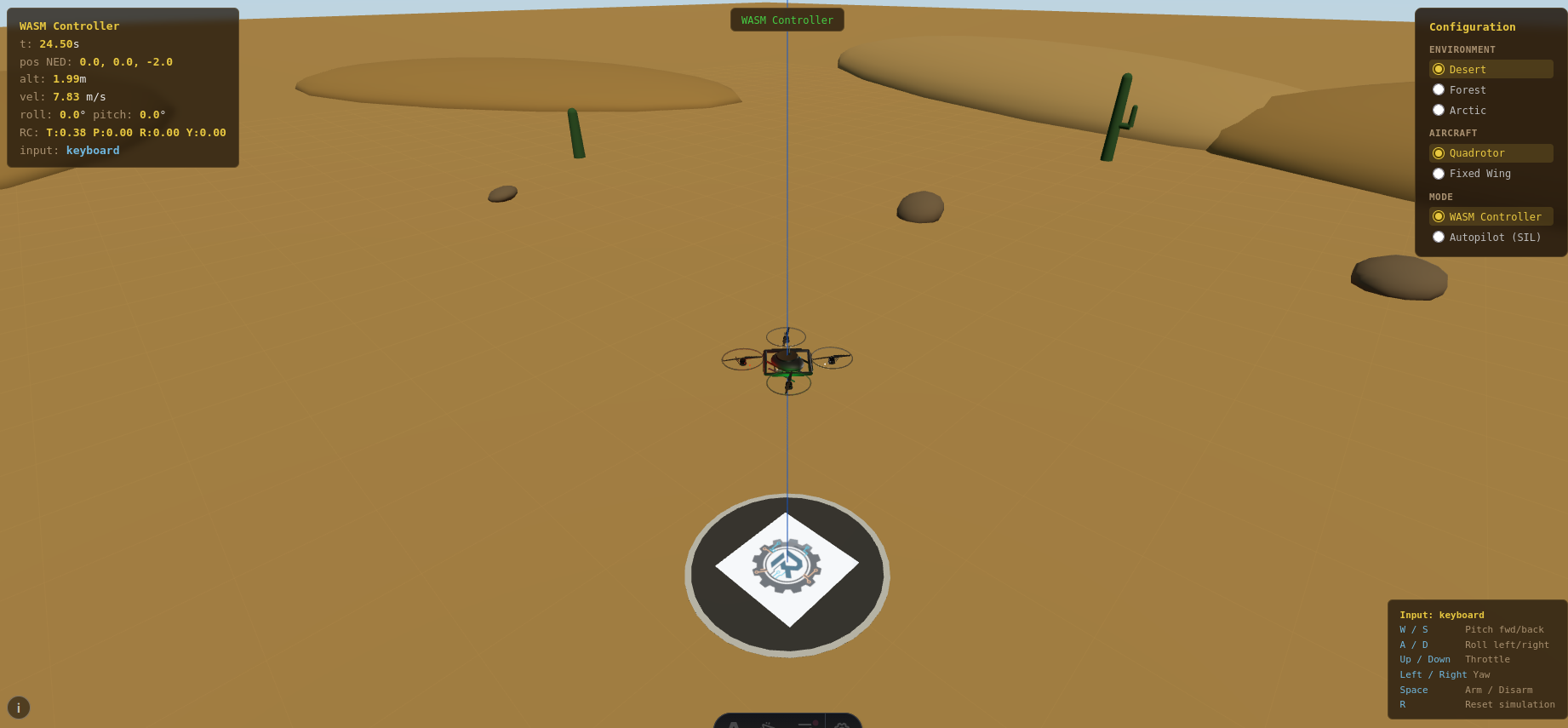}
\caption{Software-in-the-loop setup: Rumoca runtime as realtime plant for Cerebri.}
\label{fig:sil}
\end{figure}

The first case study runs Rumoca's native simulation runtime as a
realtime SIL plant for the CogniPilot Cerebri autopilot \cite{cognipilot}.
The Modelica source is a 21-state quadrotor: six-degree-of-freedom
rigid-body kinematics with a rotation-matrix attitude representation with
Euler's equations for the angular dynamics

The runtime advances the integrator in lock-step with the autopilot,
exchanging named-signal frames with Cerebri over UDP through a FlatBuffers
codec at a fixed step rate of 250 Hz. A USB gamepad feeds
the autopilot's setpoint channel, so the operator flies the simulated
vehicle through the controller. A Three.js scene renders the state in a
browser at the step rate, and a small WebSocket proxy bridges the UDP feed
for remote operators. \autoref{fig:sil} shows the setup.

The point is not realtime control quality. It is that the same compiler
producing the symbolic artifacts of \ref{sec:multibackend} also produces a
Rust binary deterministic enough to hold a step-locked loop against a
production autopilot, without giving up the equation-based modelling
surface. The only contract between the model, the SIL host, the
autopilot, and the viewer is the Modelica file and the signal schema.

\subsection{Same Source, Multiple Algebraic Backends}
\label{sec:multibackend}

The second case study drives four downstream workflows from one
hand-written Modelica source: a planar vertical-take-off-and-landing
aircraft (PVTOL), the canonical two-input textbook quadrotor. Its six states are the horizontal position
$x$, the altitude $z$, the pitch angle $\theta$, and their time
derivatives $\dot{x}$, $\dot{z}$, $\dot{\theta}$; its two inputs are the
total thrust $T$ and the pitching moment $M$; and its parameters are the
mass $m$, the moment of inertia about the pitch axis $J$, and the
gravitational acceleration $g$. \autoref{lst:pvtol} gives the complete
source. It is about twenty lines, small enough that the symbolic
deliverable of \ref{sec:SymPy} fits on the page and can be checked
against a textbook. Hover trim is $\theta=0$, $T=mg$, $M=0$.

\begin{lstlisting}[language=Modelica, caption={The complete PVTOL
Modelica source. This one file feeds all four backends of
\ref{sec:multibackend}.}, label={lst:pvtol}, basicstyle=\ttfamily\small]
model PVTOL "Planar VTOL aircraft"
  parameter Real m = 2.496   "Mass [kg]";
  parameter Real J = 0.0344  "Inertia about y [kg.m2]";
  parameter Real g = 9.80665 "Gravity [m/s2]";
  Real x, z, theta             "Position, pitch";
  Real x_dot, z_dot, theta_dot "Velocities";
  input Real T "Total thrust [N]";
  input Real M "Pitching moment [N.m]";
equation
  der(x)     = x_dot;
  der(z)     = z_dot;
  der(theta) = theta_dot;
  m*der(x_dot)     = -T*sin(theta);
  m*der(z_dot)     =  T*cos(theta) - m*g;
  J*der(theta_dot) =  M;
end PVTOL;
\end{lstlisting}

From this one source Rumoca emits four artifacts: a CasADi~MX module; a SymPy \texttt{Model} class; a JAX module
exposing the right-hand side as a function \texttt{ode\_fn}$(t, y,
\mathrm{args})$, with time $t$, state vector $y$, and an argument tuple
$\mathrm{args}$ holding the parameters and inputs; and an FMI~3.0
model-exchange FMU. Each artifact comes from one
Rumoca compilation; the consumer scripts below import it as an ordinary
module or FMU and never touch generated code.

\subsubsection{Linear MPC with CasADi}
\label{sec:casadi}

\begin{figure}[t]
  \centering
  \includegraphics[width=\columnwidth]{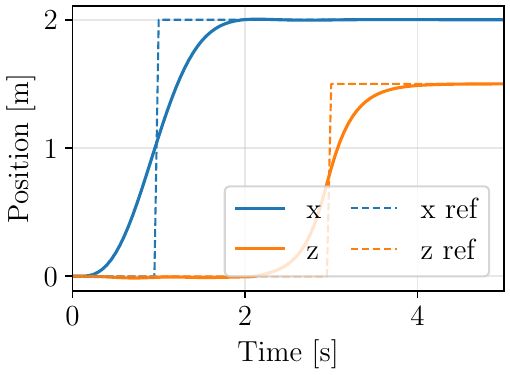}
  \caption{Linear MPC built on the Rumoca-emitted CasADi model tracking a
  step reference in the horizontal position $x$ and altitude $z$. The
  prediction model is the hover linearization; the plant is the nonlinear
  emitted model.}
  \label{fig:mpc}
\end{figure}

We differentiate the CasADi-emitted state derivative at the hover trim to
obtain the hover linearization of \ref{sec:SymPy}, and use it as the
prediction model of a linear model-predictive controller (MPC): forward
Euler at 20\,Hz over a 1.0\,s horizon, with the thrust $T$ and moment $M$
bounded, posed as a quadratic program (QP) through CasADi's \texttt{Opti}
interface and solved. The
controller runs against the nonlinear emitted plant and tracks a step
reference (\autoref{fig:mpc}), with a mean QP solve time near 10\,ms, inside
the control period. The glue code is plant-agnostic: swapping the Modelica
source re-derives the linearization automatically.

\subsubsection{Symbolic Linearization and Cross-Backend Check with SymPy}
\label{sec:SymPy}

The SymPy backend exposes the dynamics symbolically, so the Jacobians
follow in closed form. At hover, the state matrix $A$ and input matrix $B$
are
\begin{equation}
\label{eq:AB}
A = \left[\begin{matrix}
0 & 0 & 0 & 1 & 0 & 0\\
0 & 0 & 0 & 0 & 1 & 0\\
0 & 0 & 0 & 0 & 0 & 1\\
0 & 0 & -g & 0 & 0 & 0\\
0 & 0 & 0 & 0 & 0 & 0\\
0 & 0 & 0 & 0 & 0 & 0
\end{matrix}\right],
\qquad
B = \left[\begin{matrix}
0 & 0\\ 0 & 0\\ 0 & 0\\ 0 & 0\\
\tfrac{1}{m} & 0\\ 0 & \tfrac{1}{J}
\end{matrix}\right],
\end{equation}
matching the textbook PVTOL linearization, and SymPy certifies that the
controllability matrix $[\,B,\;AB,\;\dots,\;A^5B\,]$ has full rank~6
symbolically, independent of the parameters.

The same script then checks the backends against each other. We build
CasADi's analytic Jacobians from the same source, lambdify SymPy's, and
evaluate both at five random points in the state and input space. The
maximum elementwise difference is $2.8\times10^{-16}$ for $A$ and
$3.6\times10^{-14}$ for $B$. Two backends emitted from one \texttt{.mo}
file by separate templates encode numerically identical dynamics, the
guarantee a hand-ported model cannot give.

\subsubsection{Parameter Identification with JAX}
\label{sec:jax}

The JAX module exposes the dynamics as the function \texttt{ode\_fn}
introduced above, so it composes with \texttt{jit}, \texttt{grad}, and
\texttt{vmap} and slots into an adjoint-mode integrator. This makes the whole trajectory differentiable
with respect to the model's physical parameters $m$ and $J$.

\begin{figure*}[t]
  \centering
  \includegraphics[width=\textwidth]{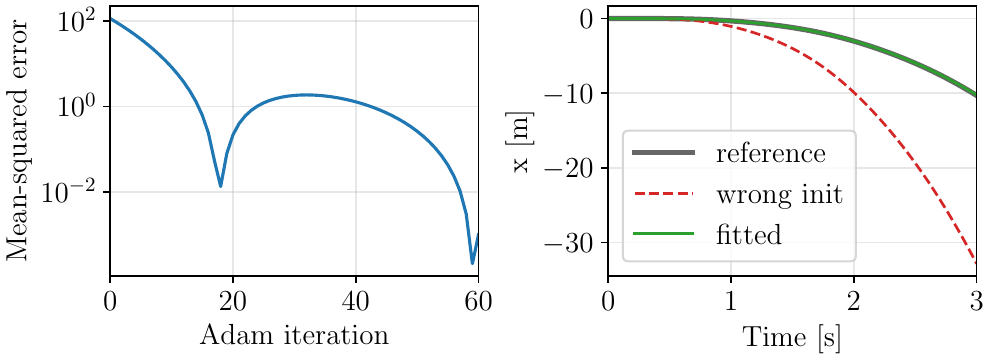}
  \caption{Parameter identification through a differentiable rollout of the
  Rumoca-emitted JAX model. Left: the mean-squared trajectory error over 60
  Adam iterations. Right: the reference, the diverging wrong-initialization,
  and the fitted trajectory in the horizontal position $x$.}
  \label{fig:paramid}
\end{figure*}

We build a reference trajectory under the ground-truth values
$m=2.496$\,kg and $J=0.0344$\,kg\,m$^2$ with a thrust-step,
sinusoidal-moment excitation over three seconds, then start from a poor
initial guess for the mass and inertia ($\hat{m}_0=0.50\,m$,
$\hat{J}_0=0.60\,J$, where $\hat{m}$ and $\hat{J}$ are the estimates) and
minimize the mean-squared trajectory error by gradient descent on
$(\log\hat{m},\log\hat{J})$. After 60 Adam iterations the recovered values
are $\hat{m}=2.4985$\,kg ($+0.10\%$) and $\hat{J}=0.034544$\,kg\,m$^2$
($+0.42\%$), with the loss falling by a factor of $5.5\times10^{5}$
(\autoref{fig:paramid}). The consumer code never reimplements the dynamics or
derives a sensitivity equation by hand; the adjoint integration comes from
the emitted module.

\subsubsection{FMI 3.0 Export}
\label{sec:fmi}

The FMI~3.0 backend produces a tool-independent FMU: a zip with the
\texttt{modelDescription.xml} interface, a compiled shared object, and the
source manifest. Loaded with \texttt{fmpy}, it holds exact hover trim
($z(2\,\text{s})=3.8\times10^{-16}$\,m) and, under a 10\% over-thrust
$T=1.10\,mg$, follows the closed-form ballistic climb to
$8\times10^{-15}$\,m. Each step costs about $25\,\mu$s at 200\,Hz, roughly
$200\times$ faster than real time, leaving headroom to host the FMU inside
a larger co-simulation rig.

Across the four workflows the modeller wrote one short Modelica source;
the consumer chose the artifact, and SymPy certified that two of the
bindings agree to machine precision.

\section{Roadmap and Discussion}\label{sec:roadmap}

Rumoca turns Modelica into a frontend: one source, authored once, emitted into
whatever ecosystem the work demands—CasADi, JAX, SymPy, FMI, or a native Rust
runtime—with cross-backend agreement to machine precision that a hand-ported model
cannot match. Two lines of work carry that thesis forward.

The first is coverage. End-to-end MSL simulation trace parity to the  OpenModelica compiler is 28\,\%, leaving room for improvement. We are taking a structured defensive coding approach to address this and focusing on built-in tracing tools that assist authors diagnose simulation and compile time errors back to source code through span preservation. The second is templates. Because code generation is template-driven, the reach of Rumoca grows by adding and hardening backends rather than rewriting the compiler. The near-term work is promoting the experimental Embedded~C target to production, broadening FMI~3.0 coverage, and bringing the JAX, Julia, and ONNX templates to the maturity of the CasADi and SymPy paths, each one extending where a single Modelica model can be deployed.

We also note that MSL parity, having served as our development and
regression corpus, functions more as a language-coverage test than as
a measure of real-world utility, which the case studies of
Section~\ref{sec:casestudies} address only from author-written models.
No Modelica collection is curated for the optimization and
machine-learning workloads Rumoca targets, so we invite the community
to contribute models that already simulate in an established compiler;
each such model is a known reference against which Rumoca's coverage
and cross-backend agreement can be measured.
Feedback from these early adopters will steer coverage priorities more
directly than MSL parity alone.

\section*{Acknowledgements}
This material is based on research sponsored by DARPA under agreement number  FA8750-24-2-0500. The  U.S. Government is authorized to reproduce and distribute reprints for Governmental purposes notwithstanding any copyright notation thereon.
\printbibliography

\end{document}